\documentclass[
aps,%
11pt,%
article,%
notitlepage,%
oneside,%
onecolumn,%
nobibnotes,%
nofootinbib,%
superscriptaddress,%
noshowpacs,%
centertags%
fleqn]%
{revtex4}

\usepackage{amssymb}
\usepackage{amsmath}
\usepackage{tabularx}

\usepackage[pdftex]{graphicx}
\usepackage{epstopdf}

\def\thesection       {\arabic{section}}
\def\p@section        {}

\def\p@subsection     {\thesection.}

\DeclareMathOperator{\Rot}{rot}

\begin{document}


\title{ON THE NATURE OF THE M{\"O}SSBAUER EFFECT}%

\author{A.V.~Kirichok}
\email{sandyrcs@gmail.com}
\author{V.M.~Kuklin}
\affiliation{V.N.~Karazin Kharkiv National University , Institute for High Technologies\\4 Svobody Sq., Kharkiv 61022, Ukraine}

\author{A.G.~Zagorodny}
\affiliation{Bogolyubov Institute for Theoretical Physics\\14-b, Metrolohichna str., Kiev, 03680, Ukraine}

\begin{abstract}
	The spectrum of electromagnetic waves emitted by an oscillator trapped in an external potential well is studied. It is assumed that the natural frequency of the oscillator is much greater than the frequency of oscillations in the potential well. We consider the quantum model of emission with regard to the recoil effect. The highest intensity of the absorption and emission lines is observed on the natural frequency of the oscillator when the recoil energy is equal to the energy of the quantum of low-frequency (LF) oscillations in the potential well. 
	
	A certain decrease in the amplitude of the emission and absorption lines is noted caused by the oscillations of the potential well due to, for example, the presence of the phonon spectrum. The relaxation times of the oscillator LF motion in the potential well, resulted from the phonon emission, are estimated. It is concluded that these processes have no influence on the observed features of the emission and absorption of high-frequency photons. This model can be applied to description of the emission and absorption of gamma rays in the crystal structures even in the presence of the phonon spectrum on condition that the relaxation time of the LF movements in the potential well is greater than the lifetime of the HF oscillator.
\end{abstract}

\pacs{42.50.Fx, 52.35.Mw}
\keywords{Mossbauer effect, recoil effect, quantum oscillator}

\maketitle

\section{Introduction}

The scattering~of high-energy photons by free electrons results in a decrease in energy of photons due to the recoil effect (the Compton effect). This fact together with the phenomenon of the photoelectric effect~confirmed the basic principles of quantum~theory of radiation \cite{born1989atomic, iwanenko1953klassische, akhiezer1965quantum}. The processes~underlying the interaction of radiation~and matter are characterized by an impressive variety and form the basis for many physical~research directions \cite{abramov1977principles,vinetsky1979radiation,wertheim2013mossbauer}. One of the problems that arises when considering the processes of absorption and emission by a matter~is the problem~of interaction~with the external radiation field~of the oscillating particle~trapped~into the potential well~formed by the spatial structure~of the medium. This problem~requires the use of~methods of quantum electrodynamics for describing the behavior of~an excited~oscillator in a potential well. 

When the energy of the single photon becomes comparable to the kinetic energy of the oscillator, the recoil effects should be taken into account. As follows from the energy and momentum conservation laws, the Lamb-Dicke parameter $kb$ \cite{javanainen1981laser}, which is proportional to both the wave number of the resonant photon $k$  and the spatial spread of the oscillator ground state wavefunction $b$, becomes in this case rather large (as will be shown below $kb\approx 2$) and extra spectrum lines appear both in the emission and absorption spectra, shifted on the value of the recoil energy $E_{r}/\hbar$. Similar effect was studied in \cite{li2013recoil}. The interaction between the external field and this extended set of spectral lines can change the nature of absorption and emission processes of the oscillator trapped in the potential well.

The purpose of this work is to analyze the interaction between a charged particle-oscillator trapped in a potential well and external electromagnetic field. We found that the highest intensity of the absorption and emission lines is observed on the natural frequency of the oscillator $\omega _{0} $  when the recoil energy $E_{r} $  is equal to the energy of low-frequency quantum  $\hbar \Omega $ ($\Omega$ is the frequency of oscillations in the potential well). We also discuss the role of relaxation of low-frequency motion in the potential well due to emission of sound waves and conditions under which the relaxation does not affect the considered emission and absorption of high-frequency quanta. 

\section{The emission of vibrating oscillator}

Following the method described in \cite{born1989atomic}, we consider the quantum model of the oscillator with a charge $e$, mass $m$ and natural frequency $\omega _{0} $, which oscillates as a whole in the potential well aligned along axis \textit{OZ}. Let the oscillator emits the electromagnetic wave in the same direction $\vec{k}=(0;\; 0;\; k_{z} )$. The components of the oscillator's velocity vector are: 
\begin{equation} \label{eqn1} 
v_{x} =v_{x0} \cos \omega _{0} t=a\omega _{0} \cos \omega _{0} t, v_{z} =b\Omega \cos (\Omega t).          
\end{equation} 
The electromagnetic field can be found from
\begin{equation} \label{eqn2)} 
\vec{E}=-\frac{1}{c} \dot{\vec{A}},\vec{H}=\Rot\vec{A}.                      
\end{equation} 
The vector potential has the components
\begin{equation} \label{eqn3} 
A_{x} =q(t)\sqrt{2} \cos (kz+\delta ), A_{y} =0, A_{z} =0.  
\end{equation} 
Phase $\delta $ depends on the oscillator orientation. The form \eqref{eqn3} choice is determined by normalizing condition, so that the integral of the vector potential squared in the unit volume equals to unity and $q(t)$ satisfies the equation 
\begin{equation} \label{eqn4} 
\ddot{q}+\omega ^{2} q=0.        
\end{equation} 
The total field energy inside \textit{V}-volume is equal to 
\begin{equation} \label{eqn5} 
U=\frac{V}{4\pi c^{2} } \frac{1}{2} (\dot{q}^{2} +\omega ^{2} q^{2} ).     
\end{equation} 

Let define the effective mass of the oscillator as
\begin{equation} \label{eqn6} 
m_{eff} =V/4\pi c^{2} .   
\end{equation} 
As follows from Eqs. \eqref{eqn1} and \eqref{eqn3} , the $x$-component of vector potential $A_{x}$ in the location point of the oscillator is equal
\begin{equation} \label{eqn7} 
A_{x} =\sqrt{2} q_{0} \cos \omega t\cdot \cos \left(kb\sin \Omega t+\delta \right) 
\end{equation} 
Suppose the particle occupies the lowest energy level in an external potential well and consider the case, when it remains inside the well after the emission or absorption of a quantum. That means the recoil energy to be insufficient for the particle to leave the potential well it stays in. 

After absorbing a high-frequency quantum $E_{\nu } =\hbar (\omega _{0} +\Omega )$, the particle-oscillator gets the recoil momentum $mV$ and begins slow oscillations inside the potential well. The energy of the quantum, emitted by the particle-oscillator with natural frequency $\omega _{0} $ is equal $E_{\nu } =\hbar (\omega _{0} -\Omega )$ due to the recoil effect.

The conservation laws for the case of quantum absorption are:
\begin{equation} \label{eqn8} 
\hbar (\omega _{0} +\Omega )/c=mV,     
\end{equation} 
\begin{equation} \label{eqn9} 
\hbar \Omega =mV^{2}/2.       
\end{equation} 

It is noticeable that the oscillation energy of the particle in the potential well equals to $\hbar \Omega $. That is why the exciting quantum energy has to exceed the oscillator`s energy $\hbar \omega _{0} $ by this value. Assuming the condition 
\begin{equation} \label{eqn10} 
\hbar \omega _{0} \ll 2mc^{2}  
\end{equation} 
is fulfilled and using Eqs. \eqref{eqn8}-\eqref{eqn9}, one can find the oscillation frequency inside the potential well
\begin{equation} \label{eqn11} 
\Omega \approx \hbar \omega ^{2} _{0} /2mc^{2} .       
\end{equation} 
It is not difficult to show that Eq. \eqref{eqn11} still stands and for the case of emission. Besides, it follows from Eqs. \eqref{eqn1}, \eqref{eqn8}, \eqref{eqn9} that 
\begin{equation} \label{eqn12} 
V=b \Omega =\hbar \omega _{0} /mc.       
\end{equation} 

The Hamiltonian of the system including the oscillator and the field can be represented as \cite{born1989atomic}
\begin{equation} \label{eqn13}
H=\frac{1}{2m} \left(\vec{p}-\frac{e}{c} \vec{A}\right)\left(\vec{p}-\frac{e}{c} \vec{A}\right)+e\Phi.
\end{equation} 
The interaction part may be written out as 
\begin{multline} \label{eqn14} 
 H'=-ev_{x} {A_{x}/c}  =-\frac{ev_{x0} }{c} q_{0} \sqrt{2} \cos \delta \exp \left( i\omega t+ikb\sin \Omega t\pm \Omega t\right)  \cos \omega _{0} t\approx \\ 
{\approx -\frac{ev_{x0} }{2c} q_{0} \sqrt{2} \cos \delta \sum _{m}J_{m}  (kb)\exp \left( i(\omega -\omega _{0m} )t\pm \Omega t\right),} 
\end{multline} 
where $v_{x} =v_{x0} \cos \omega _{0} t$. This means that the external field with frequency $\omega $ interacts with the set of oscillators, which frequencies are $\omega _{0m} =\omega _{0} +m\Omega $ and the amplitude is proportional to $J_{m} (kb)$, where $J_{m} (x)$ is the Bessel function with argument $kb\approx \omega _{0} b/c$/ In other words, the system possesses $m$energy levels the transfer on each of them can be realized independently.  The additional term in the exponent $\pm \Omega t$ which doesn't follow from the classic description, takes into account the frequency variation when the HF quantum is emitted (upper sign) or absorbed (lower sign).

Representing the system ``oscillator in the potential well'' as a set of oscillators with frequencies $\omega _{0m} =\omega _{0} +m\Omega $ and imposing the condition of temporal synchronism one can find that for the external field with the frequency
\begin{equation} \label{eqn15} 
\omega =(m\mp 1)\Omega +\omega _{0} ,   
\end{equation} 
the interaction Hamiltonian \eqref{eqn14} takes the form 
\begin{equation} \label{eqn16} 
H'=-\frac{ev_{x0} }{c} q_{0} \sqrt{2} \sum _{m}J_{m} (kb) \cos \delta .               
\end{equation} 
The upper sign in Eq. \eqref{eqn15} corresponds to the emission of the quantum with taking into account the recoil effect (the frequency of the external field at this is less than the proper frequency of the oscillator on the value of ${\Omega/2\pi}$, $m=0$) . The lower sign corresponds to the absorption (the frequency of the external field at this is greater than the proper frequency of the oscillator on the value ${\Omega/2\pi}$, $m=0$). Note that the correction to the matrix element due to the recoil effect is proportional to the small parameter $\hbar \omega _{0} /mc^{2}$  both for the individual processes of absorption and emission, as well as for the Compton scattering.

When the oscillator in rest, $b=0$, the only term in the sum \eqref{eqn16} is different from zero ($m=0$):
\begin{equation} \label{eqn17} 
H'=-\frac{ev_{x0} }{c} q_{0} \sqrt{2} \cos \delta .       
\end{equation} 
Thus, the frequency of the absorbed radiation $\omega =\omega _{0} +\Omega $ differs from the frequency of the emitted radiation $\omega =\omega _{0} -\Omega $ on the value of $2\Omega $ that corresponds to the value of the double recoil energy. 

For $b\ne 0$, the interaction Hamiltonian for the most interesting case of the absorption and emission on the proper frequency of the oscillator $\omega _{0} $ becomes: 
\begin{equation} \label{eqn18} 
H'=-\frac{e\cdot v_{x} }{c} q\sqrt{2} J_{\pm 1} (kb)\cos \delta .      
\end{equation} 

The interaction Hamiltonian for emission on the frequency $\omega =\omega _{0} -\Omega $ and absorption on the frequency $\omega =\omega _{0} +\Omega $ is similar to Eq. \eqref{eqn18}, where $J_{\pm 1} (kb)$ should be replaced by $J_{0} (kb)$. 

Let estimate the value of the argument of Bessel functions. Since the frequency of LF oscillations in the potential well is $\Omega =v/b$ and the energy of quantum $\hbar \Omega $ is equal to the recoil energy then it follows from Eq.\eqref{eqn12} \cite{kuklina2009relative}:
\begin{equation} \label{eqn19} 
kb\approx \frac{\omega _{0} b}{c} \approx 2.                             
\end{equation} 
The matrix element of the interaction Hamiltonian \eqref{eqn18} can be written as 
\begin{equation} \label{eqn20} 
H_{if} =-\frac{e}{c} \sqrt{2} \omega _{0} x_{cd} q_{nn'} J_{\pm 1} (kb)\cos \delta .      
\end{equation} 
The subscripts $c$ and $d$ indicate two states of the emitted ($n,\; n+1$) or absorbed ($n,\; n-1$) field. For the absorption case $|q_{nn'} |^{2} =|q_{n,n-1} |^{2} =n|q_{01} |^{2} $ and for the emission case $|q_{nn'} |^{2} =|q_{n+1,n} |^{2} =(n+1)|q_{01} |^{2} $. 

Taking into account the expression for effective mass \eqref{eqn6}, we can write the matrix element for the spontaneous emission \cite{born1989atomic}
\begin{equation} \label{eqn21} 
|q_{01} |^{2} =\frac{hc^{2} }{V\omega _{0} } ,           
\end{equation} 
and
\begin{equation} \label{eqn22} 
|H_{if} |^{2} =-\frac{2e^{2} }{c^{2} } \frac{hc^{2} }{V}\omega _{0} \left(x_{cd}^{2} +y_{cd}^{2} \right) J_{1}^{2} (kb)\cos ^{2} \delta \left\{\begin{array}{c} {n+1} \\ {n} \end{array}\right\} ,    
\end{equation} 
where the upper value corresponds to emission and the lower value to absorption. The transition probability can be found by taking the product of Eq. \eqref{eqn22} and $4\pi ^{2} \rho (\nu _{cd} )/h^{2} $, where $\rho (\nu _{cd} )$ is the oscillation density, with taking into account the averaging over initial phases $\langle \cos ^{2} \delta \rangle =1/2$
\begin{equation} \label{eqn23} 
P_{if} =\frac{4\pi ^{2} }{h^{2} } |{\rm H} _{if} |^{2} \rho =\frac{8\pi e^{2} }{hc^{3} } \omega _{0}^{2} \left(|x_{cd} |^{2} +|y_{cd} |^{2} \right)J_{1}^{2} (kb)\cos ^{2} \delta \left\{\begin{array}{c} {n+1} \\ {n} \end{array}\right\}.    
\end{equation} 

Note that the probability of absorption on the frequency $\omega _{0} +\Omega $ and the probability of emission on the frequency $\omega _{0} -\Omega $ can be obtained by replacing $J_{1}^{2} (kb)$ by $J_{0}^{2} (kb)$ in Eq. \eqref{eqn23}.  Multiplying Eq.\eqref{eqn23} by $\hbar \omega _{0} $ one can obtain the intensity of radiation along axis \textit{OZ}  and integrating over the angle $\theta =\vec{k}\wedge O\vec{Z}$ -- the total intensity over all directions. 

It is easy to verify that since  $J_{1}^{2} (kb)\gg J_{0}^{2} (kb)$ the intensity of absorption and emission spectral lines on the proper frequency of the oscillator $\omega _{0} $ exceeds the intensity on the frequency $\omega _{0} \pm \Omega $ by an order. Note that the width of the potential well should be greater than $b$ and its depth should exceed the recoil energy. The proposed quantum-mechanical description may be more correctly clarify the mechanism of emission and absorption of $\gamma $- quanta without recoil [6]. In addition, it can be shown that taking into account fast oscillations of the potential well as whole with a frequency $\omega _{S} $ and amplitude $b_{S} $, which satisfy the conditions $b_{S}^{2} \omega _{S}^{2} >b^{2} \Omega ^{2} $ and $b_{S}^{2} \ll b^{2} $,  results in  decrease of the vector potential $A_{x} $ and interaction energy ${\rm H} '$, defined by Eqs. \eqref{eqn12} and \eqref{eqn16}, by the factor $\exp (-W/2)=(1-b_{s}^{2} /b^{2} )$ and reduces the transition probability \eqref{eqn23} in $\exp (-W)$ times. 

In a similar way, the presence of a relatively high frequency phonon spectrum in the environment will effect on the probability of the transition. For the case of a rather broad spectrum, this influence can be described by a  factor $\exp \{ -W\} =(1-b^{-2} \sum _{i=1}b_{i}^{2}  )$ under the conditions $b^{2} \Omega ^{2} \ll \sum _{i=1}b_{i}^{2} \omega _{i}^{2}  $ and $b^{2} >\sum _{i=1}b_{i}^{2}  $.

\section{The relaxation of low-frequency oscillations}

When the above-discussed system is placed in solid medium, it is necessary to consider a possibility of emission of a low-frequency phonon $\Omega =\omega _{0} (\hbar \omega _{0} /2m c^{2} )$. The relatively low velocity acquired by an oscillator due to recoil is often significantly less than the phase velocity of phonons
\begin{equation} \label{eqn24} 
v_{s} \gg c(\hbar \omega _{0} /2m c^{2} ).                       
\end{equation} 
This makes impossible the direct transfer of kinetic energy to phonon. This is evidenced by the inability to fulfill the requirements of conservation of energy and momentum. For example, it can be shown for the momentum that 
\begin{equation} \label{eqn25} 
\hbar \omega _{0} /c \gg \hbar \Omega /v_{s} =\hbar \omega _{0} \left(\frac{\hbar \omega _{0} }{2mc^{2} } \right)/ v_{s}  =\frac{\hbar \omega _{0} }{c} \frac{\hbar \omega _{0} }{2mc^{2} } \frac{c}{v_{s} } .    
\end{equation} 
However, if the oscillator is trapped to the potential well, its movement becomes irregular and it becomes capable to emit phonons. At this, the spatial period of low-frequency oscillations in the potential well
\begin{equation} \label{eqn26} 
b=\frac{V }{\Omega } =\frac{\hbar \omega _{0} }{m c} \frac{2m c^{2} }{\hbar \omega _{0} ^{2} } =\frac{c}{\omega _{0} }  
\end{equation} 
is much less than the wavelength of phonon oscillations, which frequency is equal to the frequency of oscillations in the potential well
\begin{equation} \label{eqn27} 
k_{s} b=\frac{\Omega }{v_{s} } \frac{c}{\omega _{0} } =\frac{\hbar \omega _{0} }{2mc^{2} } \frac{c}{v_{s} } \ll 1.                         
\end{equation} 

Let estimate the lifetime of the low-frequency oscillator. If the lifetime occurs much greater than the period of oscillations, the above consideration of the isolated system "oscillator in potential well" remains applicable for the case when this system is placed in a medium. In other words, the relaxation processes caused by the interaction with the phonon spectrum can be neglected. 

In one-dimensional case, the perturbation of the medium density generated by low-frequency motion of the oscillator can be written as 
\begin{equation} \label{eqn28} 
\rho _{ext} (t,x)=m \delta [x-V \cos (\Omega t)] 
\end{equation} 
Its Fourier image is equal
\begin{equation} \label{eqn29} 
\rho (\omega ,k)=\frac{\omega ^{2} }{v_{s}^{2} k^{2} -\omega ^{2} } \frac{m}{4\pi } \frac{kb}{2} [\delta (\omega -\Omega )-\delta (\omega +\Omega )] 
\end{equation} 
The inverse transform gives
\begin{equation} \label{eqn30} 
\rho (x,t)=-\frac{mb\Omega ^{2} }{8v_{s}^{2} } [\sin (\Omega t-k_{0} z)+\sin (\Omega t+k_{0} z)].    
\end{equation} 
Since the perturbation of the medium speed $u=v_{s} \rho /\rho _{0} $, the ratio of emission intensity to the energy of the quantum is equal 
\begin{equation} \label{eqn31} 
\frac{I}{\hbar \Omega} =\frac{1}{16} \frac{m}{\rho _{0} \lambda _{s} } \Omega .                           
\end{equation} 
The small parameter ${m/\rho _{0} \lambda _{s} }$ here is equal to the ratio of oscillator's mass to the total mass of similar particles located on the wavelength of sound wave.  The relaxation time of low-frequency oscillations 
\begin{equation} \label{eqn32} 
\tau _{LF} \approx 8\rho _{0} \lambda _{s} /\pi m\Omega  
\end{equation} 
occurs, as was supposed earlier, much greater than the period of low-frequency oscillations in the potential well. Moreover, if $\tau _{LF} $ far exceeds the lifetime of high-frequency quantum, the relaxation process does not affect the character of emission and absorption of photons considered above. Note, that in the three-dimensional case the characteristic relaxation time of low-frequency motion 
\begin{equation} \label{eqn33} 
\tau _{LF} \approx 3(\rho _{0} \lambda _{s}^{3} /m)(\omega _{0} /\pi ^{2} \Omega _{0}^{2} ) 
\end{equation} 
is proportional to a large parameter $\rho _{0} \lambda _{s}^{3} /m$. This parameter is equal to the ratio of total mass of atoms within a cube with edge $\lambda $ to the mass of single atom. The factor $(\omega _{0} /\Omega )$ is large too. In this case the lifetime of high-frequency oscillator may occur less than the relaxation time of low-frequency motion caused by emission of sound.

\section{Conclusion}

We considered the quantum-mechanical model of electromagnetic emission by one-dimensional oscillator, trapped in the external potential well. On the assumption of the energy of the emitted quantum is much less that the rest energy of the oscillator (${\hbar \omega _{0} \mathord{\left/ {\vphantom {\hbar \omega _{0}  mc^{2} }} \right. \kern-\nulldelimiterspace} mc^{2} } \ll 1$), the perturbation of the Hamiltonian matrix elements caused by the recoil effect can be neglected, that simplifies their calculation. 

It is shown that the intensity of the absorption and emission lines at the natural frequency $\omega _{0} $ of the oscillator significantly exceeds the intensity of other spectral lines, particularly at frequencies $\omega _{0} \pm \Omega $. This is caused by the equality of the oscillation energy in the potential well to the recoil energy, and the fact that the amplitude of the oscillation in the potential well in this case is comparable to the wavelength of the radiation. 

It is noted a slight decrease in the amplitude of the emission and absorption lines due to the oscillations of the potential well as whole, due to the presence of the phonon spectrum. It is estimated the relaxation time of low-frequency movements in the potential well caused by the phonon radiation to the environment. It is concluded that these processes have no effect on the observed features of the processes of emission and absorption of high-frequency photons.

Such a quantum-mechanical description may be more correctly explains the phenomenon of emission and absorption of electromagnetic field quanta without recoil by the matter [6]. Really, the emission of the oscillator with natural frequency $\omega _{0} $ produces the equidistant spectrum $\omega _{0} +n\, \Omega $ that can be considered as result of emission of a set of oscillators with oscillation amplitude proportional to $J_{n}^{2} (kb)$. Note that the existence of such an equidistant spectrum of the oscillator captured in the potential well was also discussed in \cite{li2013recoil} under slightly different conditions. Moreover, since the amplitude of low-frequency oscillations in the potential well $b$ is comparable with the emission wavelength $\lambda ={2\pi \mathord{\left/ {\vphantom {2\pi  k}} \right. \kern-\nulldelimiterspace} k} $, the emission and absorption on the proper frequency $\omega _{0} $ dominate in this case. The nature of the high-frequency oscillator is not critical for the manifestation of the above-described properties of the system.

Note, that the application of this model to description of emission and absorption of $\gamma $-quanta in crystal structures is possible even in presence of the phonon spectrum, when the relaxation time of low-frequency motions in the potential well significantly exceeds the period of oscillations in the potential well and the lifetime of high-frequency oscillator. For example, the absorption of $\gamma $-quanta with energy 14,4 kEv and 23,8 kEv by nuclei ${}^{57}Fe$  and ${}^{119}Sn$ correspondingly results in accordance with formula $\omega _{0} b/c \approx 2$ to the oscillations of atoms in the potential well, formed by the crystal, with amplitude excursion $0,55\cdot 10^{-8} $ cm and $0,33\cdot 10^{-8} $cm. The relaxation time of this oscillatory motion is about 0.1 sec for ${}^{57}$Fe  and about 0.01 sec for ${}^{119}$Sn, that is much more than life-time of excitation state of these nuclei. The absorption and emission on the proper frequency of the atomic nucleus $\omega _{0} $ is dictated basically by the presence of the sufficient number of atoms, which oscillate in the potential wells formed by the crystal \cite{kuklina2009relative}. 

\section{References}

\bibliography{mossbauer}

\end{document}